\documentclass[letterpaper,10pt,twocolumn]{article}

\usepackage{geometry}
\usepackage{graphicx}
\usepackage{amsmath,amssymb}
\usepackage{subfigure,multirow}
\usepackage{csquotes}
\usepackage{authblk}
\usepackage{url}

\renewenvironment{abstract}{%
\hfill\begin{minipage}{\textwidth}\leftskip0.5in \rightskip0.5in
}{
\par\noindent\end{minipage}\vskip 1.5em \par}

\makeatletter
\renewcommand\@maketitle{%
\centering
\begin{minipage}{\textwidth}
\let\footnote\thanks 
{\centering\LARGE \@title \par }
\vskip 1.5em
{\centering\large \@author \par}
\end{minipage}
\vskip 1em \par
}
\makeatother

\let\OLDthebibliography\thebibliography
\renewcommand\thebibliography[1]{
  \OLDthebibliography{#1}
  \setlength{\parskip}{0pt}
  \setlength{\itemsep}{0pt plus 0.3ex}
}
\title{\textbf{Integrated Silicon Photonic Transmitter for Polarization-Encoded Quantum Key Distribution}}

\author[1,$\dagger$,*]{Chaoxuan Ma}
\author[1,3,$\dagger$]{Wesley D. Sacher}
\author[2,$\dagger$]{Zhiyuan Tang}
\author[1]{Jared C. Mikkelsen}
\author[1]{Yisu Yang}
\author[1,4]{Feihu Xu}
\author[1,2]{Hoi-Kwong Lo}
\author[1,*]{Joyce K. S. Poon}

\affil[1]{Department of Electrical and Computer Engineering, University of Toronto, 10 King's College Road, Toronto, Ontario, M5S 3G4, Canada}
\affil[2]{Department of Physics, University of Toronto, 60 St. George St., Toronto, Ontario, M5S 1A7, Canada}
\affil[3]{Departments of Physics \& Applied Physics, California Institute of Technology, 1200 E. California Blvd., Pasadena, California 91125, USA}
\affil[4]{Research Laboratory of Electronics, Massachusetts Institute of Technology, 77 Massachusetts Ave., Cambridge, Massachusetts 02139, USA}

\affil[$\dagger$]{These authors contributed equally to this work.}
\affil[*]{Corresponding authors: chaoxuan.ma@mail.utoronto.ca, joyce.poon@utoronto.ca}

\begin{document}

\twocolumn[\begin{@twocolumnfalse}
\maketitle  
\thispagestyle{empty}
\begin{abstract}
We present a silicon optical transmitter for polarization-encoded  quantum key distribution (QKD).  The chip was fabricated in a standard silicon photonic foundry process and integrated a pulse generator, intensity modulator, variable optical attenuator, and polarization modulator in a 1.3 mm $\times$ 3 mm die area. The devices in the photonic circuit meet the requirements for QKD. The transmitter was used in a proof-of-concept demonstration of the BB84 QKD protocol over a 5 km long fiber link.
\end{abstract}
\end{@twocolumnfalse}
]

%\maketitle must follow title, authors, abstract, \pacs, and \keywords

Leveraging the infrastructure for complementary metal-oxide-semiconductor (CMOS) electronics manufacturing, silicon (Si) photonics is emerging as a key technology for next-generation computing and communication systems with low power consumption and potentially low-cost optoelectronic integration. Many Si photonic devices for classical optical communication have been successfully demonstrated, including waveguides, high-speed optical modulators, photodetectors, wavelength converters and polarization management components \cite{soref2006past, LimJSTQE2014, DoerrFIP2015}. The availability of these building blocks,  coupled with  photon sources in Si and hybrid integration methods for single-photon detectors  \cite{bonneau2016silicon},   paves the way for Si photonics to be applied to quantum information.  Quantum photonics has traditionally relied on bulk optical devices or exquisitely fabricated, singular microphotonic devices that have limited scalability and integration compatibility with other classical computing or communication components.  Maturing Si photonic foundry fabrication services can enable Si photonic integrated circuits (PICs) for quantum information \cite{LimJSTQE2014,HochbergISSCM2013}, reducing the cost of incorporating quantum photonic functionality into classical systems.  

A  quantum technology prime for widespread use is quantum key distribution (QKD) \cite{bennett1984quantum,ekert1991quantum,lo2014secure},
which exploits  statistics of single photons  to generate secure encryption keys.  Owing to the quantum no-cloning theorem \cite{wootters1982single,dieks1982communication}, inevitable disturbances introduced by eavesdropping leads to a higher than expected quantum bit error rate (QBER) that can be detected. Moreover, unconditional security with faint laser pulses can be achieved using decoy states \cite{hwang2003quantum, lo2005decoy, ma2005practical, wang2005beating}. Most of today's commercial QKD systems use discrete components \cite{idq2016clavis}.

For chip-scale devices, silica planar lightwave circuits have been used for on-chip interferometers and waveguides \cite{nambu2008quantum, kristensen2014quantum, melen2016integrated}; however, the circuits were entirely passive.  Recently,  phase-shift and time-bin encoded 
QKD protocols have been demonstrated using transmitter and receiver PICs (without the single photon detectors) fabricated in monolithic indium phosphide (InP) and silicon nitride (SiN) integrated photonic platforms available through foundries \cite{sibson2015chip}.

In this letter, we report the first integrated transmitter PIC for  QKD  fabricated in a standard foundry Si photonic platform.   The large wafer sizes available in Si photonic foundry processes (8'' or 12'' diameters \cite{LimJSTQE2014, epiXfab2015imec, BoeufOFC2015} vs. 3''  for InP \cite{smit2014introduction} and  or 4'' for SiN \cite{heideman2009large}) and dense integration are conducive to scaling to high volume manufacturing. The transmitter PIC supports polarization-encoded QKD protocols and contains ring modulators, a variable optical attenuator (VOA) and polarization modulator. A secret key rate of 0.952 kbps was achieved with a QBER of 5.4\% for a proof-of-concept QKD demonstration using the BB84 protocol.  Phase and time-bin encoding should also be possible using Si photonics,  but such implementations  require interferometers with long on-chip delay lines. Even though polarization states are not maintained in standard single mode telecommunication fibers, the relations between the polarization states are preserved at the receiver, and the polarization states can be recovered \cite{peng2007experimental,tang2014experimental}. Polarization multiplexed protocols are becoming standard in classical optical communication \cite{WinzerJLT2010}. Polarization-encoded QKD is also preferred for free-space links, such as satellite-to-ground communications \cite{schmitt2007experimental,Meyer-ScottPRA2011}.

\begin{figure}[!t]
\subfigure[]{\includegraphics[trim= 0in 2in 0in 1in, clip, width=3.2in]{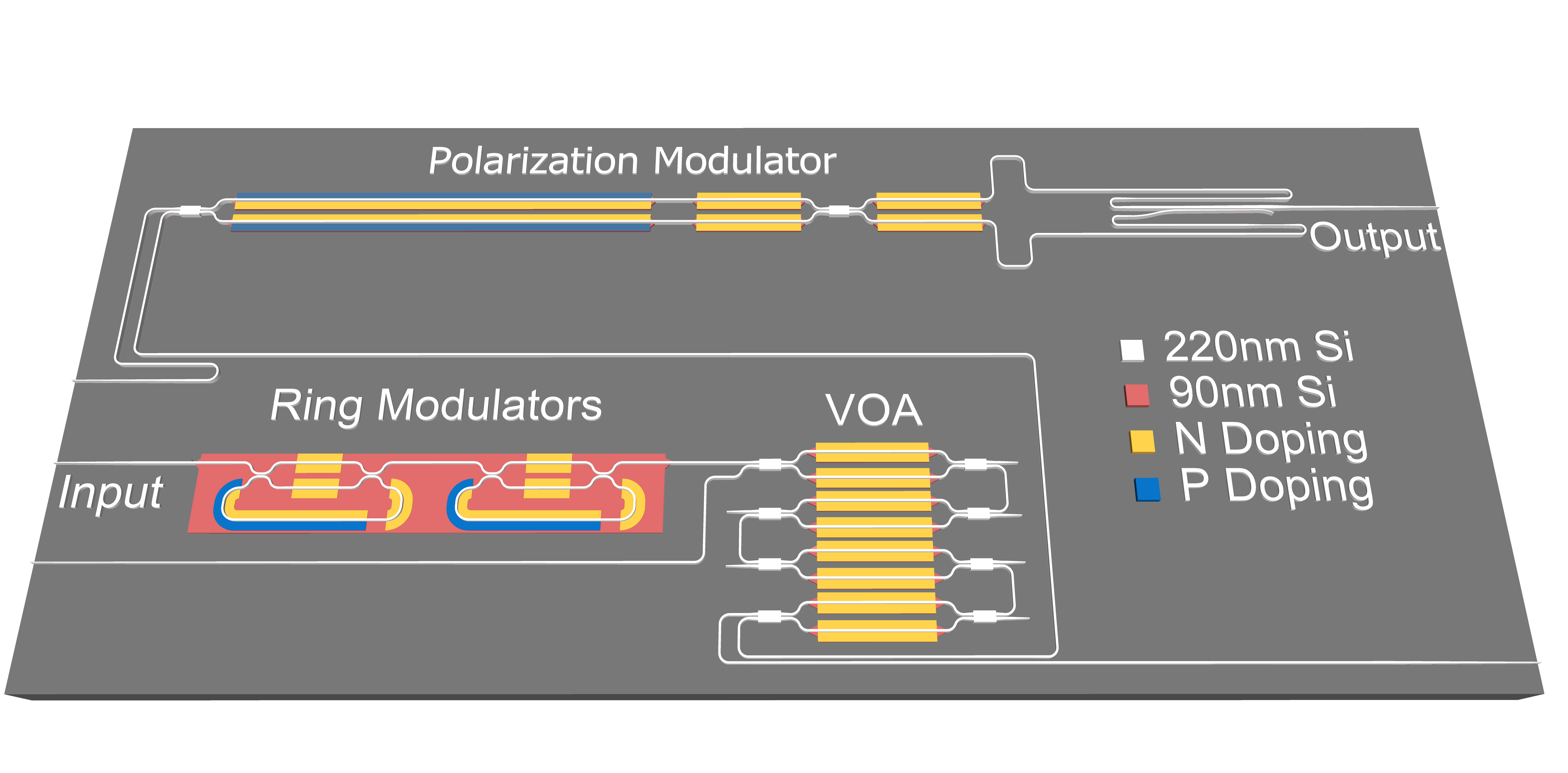}}\label{fig:schematic}
\subfigure[]{\includegraphics[trim= 0in 0in 0in 0in, clip, width=3.25in]{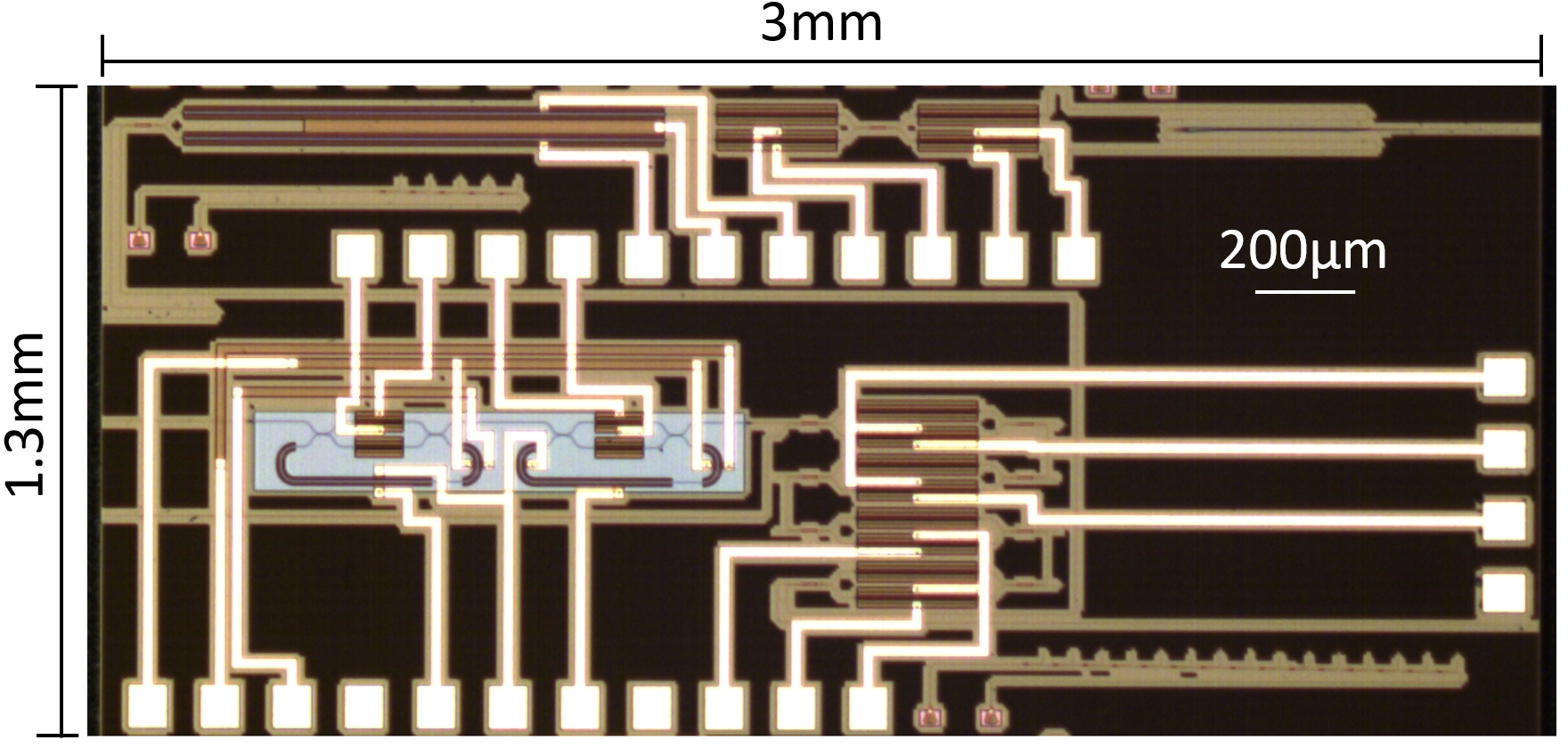}}\label{fig:micrograph}
\caption{(a) Schematic of the Si PIC transmitter for polarization-encoded QKD. (b) Optical micrograph of the  chip. %The size of the transmitter PIC is  1.3 mm $\times$ 3 mm.
}\label{fig:chip}
\end{figure}

The schematic and optical micrograph of the PIC are shown in  Fig.~\ref{fig:chip}. The transmitter, with a small size of 1.3  $\times$ 3 mm$^2$, was fabricated in the A*STAR IME baseline Si photonics process. It integrates together two identical microrings, a VOA and a polarization modulator. The first microring generates periodic nanosecond (ns) pulse trains, while the second modulates pulse intensities to create decoy and signal states, if needed.  The VOA attenuates the pulses into single photon level, and the polarization modulator prepares the polarization state of photons. Light is coupled into/out of the chip using on-chip adiabatic taper waveguide couplers and lensed fibers with a 2.5 $\mu$m spot diameter. The tip of the edge couplers has a cross section of 200 nm $\times$ 220 nm, to minimize the polarization dependent loss (PDL). Extra input and output ports are available in the PIC to enable the characterization of the individual devices, which we next describe. 

%\section{Experiemental results}

%%%%%%%%%%%%%%%%%%%%%%%%%%%%%%%%%%%%%%% 
% RINGS
%%%%%%%%%%%%%%%%%%%%%%%%%%%%%%%%%%%%%%%
\begin{figure}[!t]
\subfigure[]{\label{fig:ring_micrograph}\includegraphics[width=1.6in]{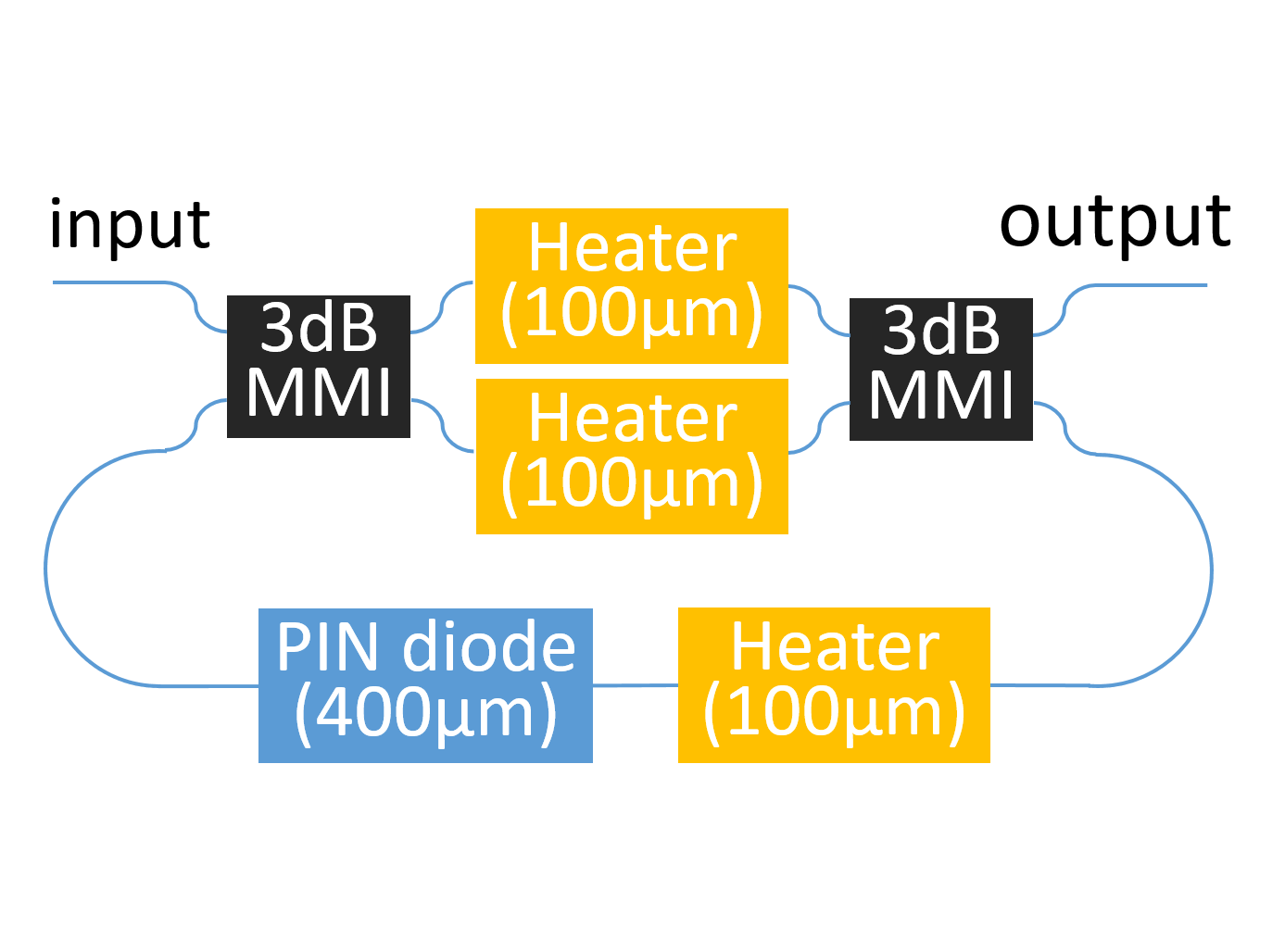}}
\subfigure[]{\label{fig:ring_spec}\includegraphics[width=1.6in]{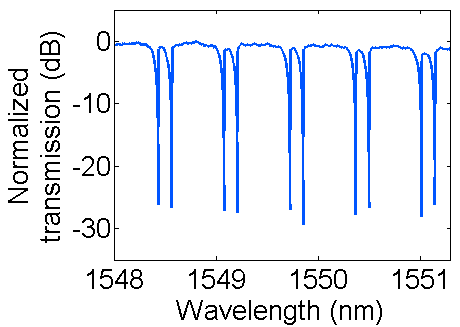}}\\
\subfigure[]{\label{fig:ring_tunability}\includegraphics[width=1.6in]{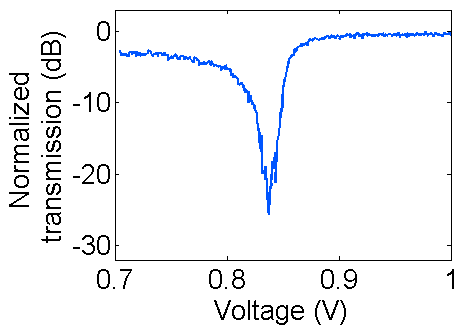}}
\subfigure[]{\label{fig:ring_pulse}\includegraphics[width=1.6in]{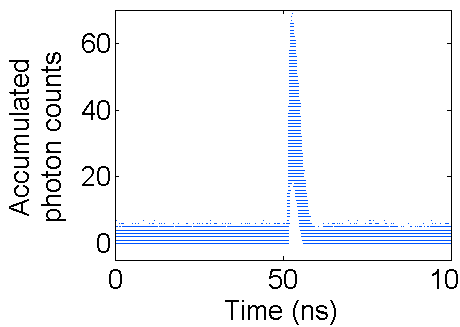}}\\
\subfigure[]{\label{fig:ring_hist}\includegraphics[width=1.6in]{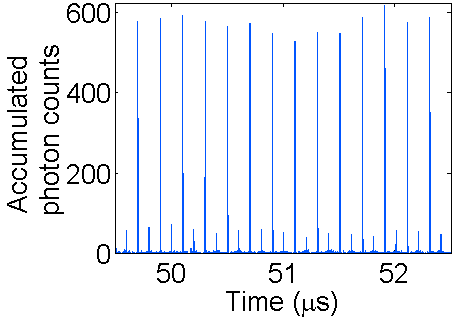}}
\subfigure[]{\label{fig:ring_dist}\includegraphics[width=1.6in]{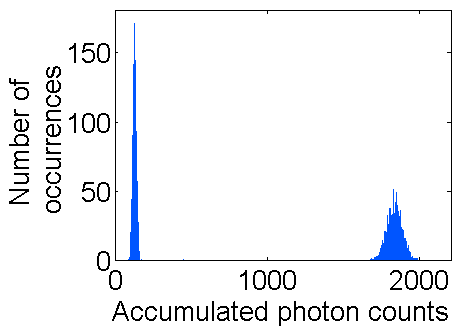}}

\caption{\label{fig:ring} (a) Schematic of the microring. (b) The normalized transmission spectrum with both microrings biased at critical coupling. (c) DC tuning of the  transmission spectrum. (d) Histogram of 8000 overlapping pulses generated by the first microring. (e) The time-dependent accumulated photon counts due to two alternating intensities from the modulation of the two microrings. The time bin is 1 ns. (f) The photon number distribution of the pulses in (e). The  photon counts were calculated using a time bin of 5 ns.}
\end{figure}

Figure ~\ref{fig:ring_micrograph} shows the schematic of the  microring modulator.  Each microring contains a 400 $\mu\mathrm{ m} $ long PIN diode phase-shifter inside the ring for modulation. A $2\times2$ Mach-Zehnder interferometer (MZI) coupler and an intracavity 100 $\mu$m long doped Si resistive heater provide independent tuning of the coupling coefficient and resonance wavelength, respectively, to achieve modulation with a high extinction ratio (ER). Thin film heaters were not available in this foundry process. Fig.~\ref{fig:ring_spec} shows the static transmission of the transmitter when both microrings were set to the critical coupling condition and slightly detuned from each other. The microrings had a free spectral range of 0.65 nm, and the minimum transmission was about -27 dB. The loaded $Q$ factor of the microrings was about $9.5 \times 10^5$.  Fig.~\ref{fig:ring_tunability} shows the tuning of the transmission at a fixed wavelength of 1549.9 nm as a DC voltage was applied to the intracavity PIN diode of one of the microrings.  A static ER of 25.6 dB was achieved by an applied voltage of only 50 mV. Fig.~\ref{fig:ring_pulse} shows the pulse shape generated by the first microring, which is a histogram of 8000 overlapping pulses. The microring was driven by a programmable pattern generator with bursts of 8000 1 ns-wide long pulses at a repetition rate of 10 MHz to generate a train of nominally identical pulses. The tests lasted for > 5 min. The optical pulses had a full width half maximum (FWHM) of 2.4 ns and the jitter was 0.9 ns FWHM. The dynamic ER  was 20 dB, which, to our knowledge, is  the highest of any Si microring and MZI modulator. A high ER  reduces the QBER penalty \cite{gisin2002quantum}. 

The two microrings, when tuned to have matching resonances, could generate pulses with varying intensity levels, which can be used for signal and decoy states. Fig. \ref{fig:ring_hist} shows the cumulative photon count of $3.75\times 10^{5}$ repetitions of a modulation pattern with alternating intensities.  The first microring was driven with 1 ns long pulses at a repetition rate of 10 MHz, and the second microring modulator was driven by a 5 MHz square-wave from an arbitrary function generator (AFG).  The voltage levels of the AFG produced the two intensities shown in Fig.~\ref{fig:ring_hist}. Fig. \ref{fig:ring_dist} shows the distribution of cumulative photon counts.  The mean photon numbers per pulse of the two states were $0.129\pm 0.003$ and $0.009\pm 0.001$, respectively.

%%%%%%%%%%%%%%%%%%%%%%%%%%%%%%%%%%%%%%% 
% VARIABLE OPTICAL ATTENUATOR
%%%%%%%%%%%%%%%%%%%%%%%%%%%%%%%%%%%%%%%
\begin{figure}[!t]
\subfigure[]{\label{fig:voa_schematic}\includegraphics[trim= 0in 0in 0in 0in, clip, width=1.2in]{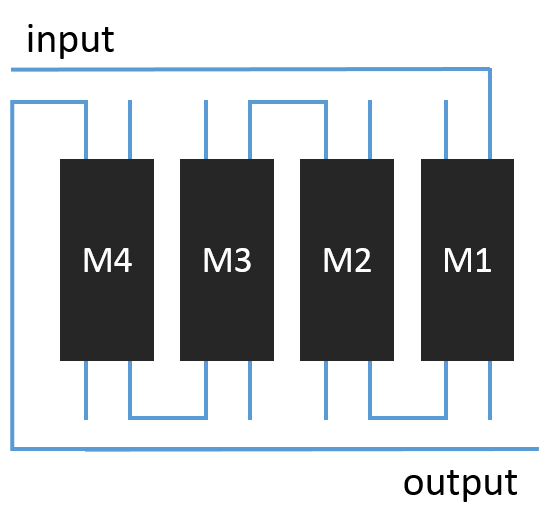}}
\subfigure[]{\label{fig:voa_tunability}\includegraphics[trim= 0in 0in 0in 0in, clip, width=2in]{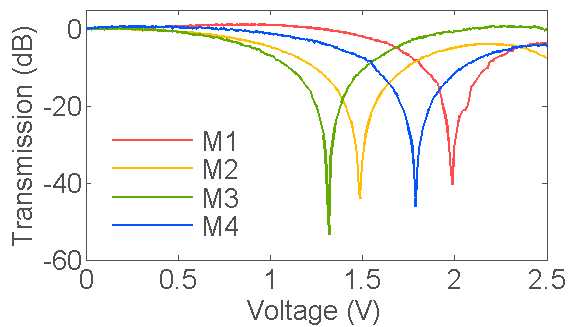}}
\caption{\label{fig:VOA} (a) Schematic of the VOA. The VOA is a cascade of four identical MZIs (labeled M1 to M4). (b) Normalized transmission of each MZI at 1550 nm. The maximum attenuation provided by each MZI is 40.3, 44.0, 53.3 and 46.4 dB. The transmission is normalized to the starting point (zero bias).}\label{fig:voa}
\end{figure}

The next device in the PIC is the VOA, which consists of four cascaded nominally balanced 2 $\times$ 2 MZIs.  The MZI has two 3 dB multimode interference (MMI) couplers separated by 300 $\mu$m  long phase-shifters each containing  a 250 $\mu$m long  heater.  The attenuation is tuned by the differential phase-shifts in the MZIs.The tuning range of each MZI at 1550 nm (labeled M1 to M4) is shown in Fig.~\ref{fig:voa_tunability}. Each MZI could provide $> 40$ dB of attenuation, so the VOA could generate an attenuation $>$160 dB.

%%%%%%%%%%%%%%%%%%%%%%%%%%%%%%%%%%%%%%% 
% POLARIZATION MODULATOR
%%%%%%%%%%%%%%%%%%%%%%%%%%%%%%%%%%%%%%%
\begin{figure}[!tbp]
\subfigure[]{\label{fig:pol_schematic}\includegraphics[width=3.2in]{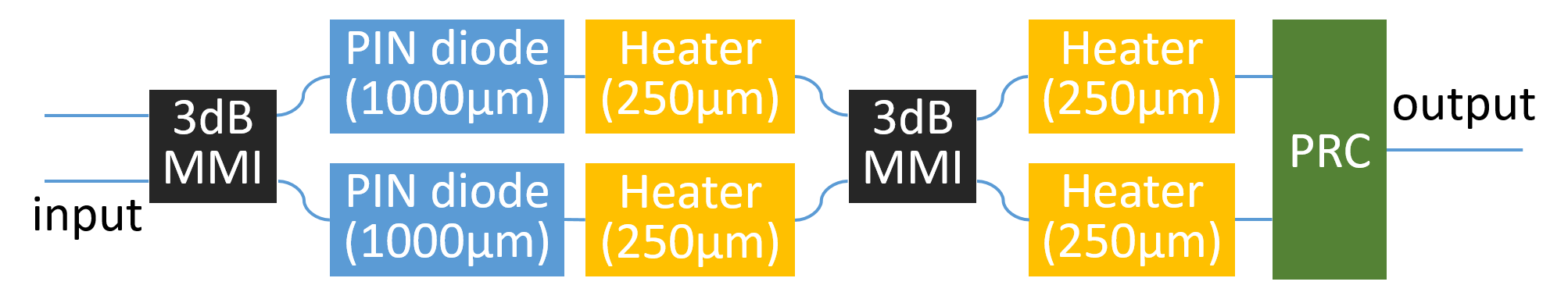}}
\subfigure[]{\label{fig:pol_transmission}\includegraphics[width=1.6in]{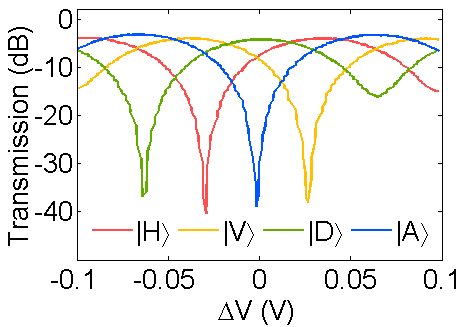}}
\subfigure[]{\label{fig:pol_extinction}\includegraphics[width=1.75in]{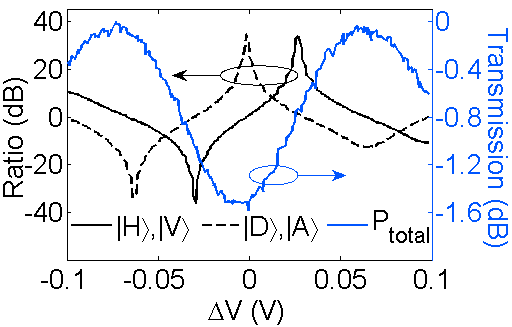}}
\caption{\label{fig:PolExp} (a) Schematic of the polarization modulator. (b) Normalized transmission of the four polarizations vs. voltage, $\pm \Delta V$, applied to the PIN diodes in the MZI at a bias of 1.2 V. (c) The total transmitted power (Normalized) and power ratio within each basis vs. $\Delta V$. }
\centering
\end{figure}

\begin{figure}[t]
\includegraphics[trim= 0in 0in 0in 0in, clip, width=3.5in]{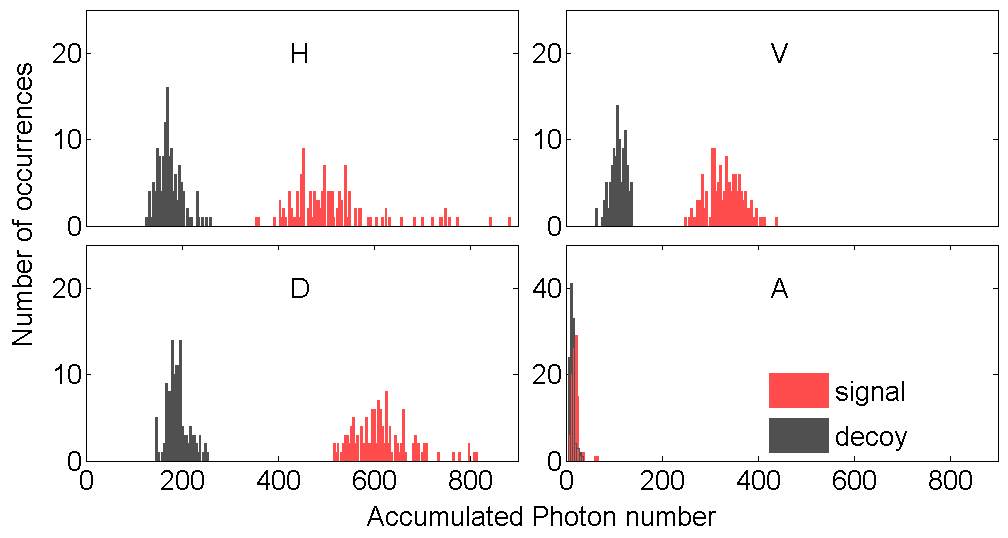}
\caption{\label{fig:pol_decoy} Projections of the polarization states onto $\vert D\rangle$ showing multi-level modulation using the second microring, VOA, and polarization modulator.}
\end{figure}

The output of the VOA connects to the polarization modulator. The polarization modulator is used to prepare the four states, \{$\vert H\rangle$, $\vert V\rangle$, $\vert D\rangle$, $\vert A\rangle$\}, that constitute two conjugate bases, where $\vert D\rangle=\frac{1}{\sqrt{2}} (\vert H\rangle+\vert V\rangle)$ and $\vert A\rangle=\frac{1}{\sqrt{2}} (\vert H\rangle-\vert V\rangle)$. In the polarization-encoding scheme, the states are identified as the polarization angles \{$0^{\circ}, 90^{\circ}, 45^{\circ}, -45^{\circ}$\}.  Fig.~\ref{fig:pol_schematic} shows the schematic of the polarization modulator, and the design is based on our  work in \cite{sacher2014polarization}. The polarization modulator consists  of a nominally balanced MZI followed by a polarization rotator combiner (PRC). The Si PIC preferentially supports the fundamental transverse electric mode (TE0).  In the polarization modulator, the MZI  controls the amplitude ratio between the two inputs to the PRC.  Then the PRC converts the TE0 mode in one of its inputs (e.g., the lower branch) into the fundamental transverse magnetic (TM0) mode, which is combined with the TE0 light from the other input (e.g., the top branch) at the output.

The MZI has a 1000 $\mu$m long PIN diode and a 250 $\mu$m long resistive heater in each phase-shifter. An additional set of heaters in the inputs to the PRC enables additional phase tuning between the TE and TM components. The PIN diodes are driven in push-pull and modulate the effective index of the Si waveguides by the plasma dispersion effect \cite{soref1987electrooptical}. Ideally, only the real part of the refractive index is modified, but the plasma dispersion effect also slightly alters the absorption coefficient, which causes a slight PDL. PDL can also be caused by fabrication errors in the PRC and fiber-to-chip coupling. The PDL at the chip output can be compensated by adding tunable attenuators at the inputs of the PRC for fine balancing of the loss between the TE and TM components. The polarization extinction ratio (PER) can be increased by including PIN diode phase-shifters at the inputs of PRC for the dynamic compensation of the phase mismatch between the two polarization components.  The measured PDL and PER are described next.

To measure the polarization tunability, the output of the polarization modulator was passed through a fused fiber polarization independent 3 dB splitter, and each output branch passed through an in-fiber polarization controller (PC) followed by a fiber-based polarization beam splitter (PBS).  The PCs were set such that each PBS was aligned to the rectilinear and diagonal basis, respectively.   Fig. \ref{fig:pol_transmission} shows the transmission for 1550 nm at  the outputs of each PBS vs. a small-signal voltage sweep, $\Delta V$, applied in push-pull mode to the PIN diodes in the on-chip polarization controller at a bias of 1.2 V. Fig. \ref{fig:pol_extinction} shows the total transmitted power and power ratio of the two orthogonal components within each basis. The PER was $> 30$ dB, and the power variation across the four polarization states was 0.9 dB. 

To show the combined functionality of the intensity modulation, attenuation, and polarization preparation, Fig. \ref{fig:pol_decoy} shows the photon number histograms of the four polarizations are projected onto $\vert D\rangle$ using  the second microring modulator, VOA, and polarization modulator.  The input to the chip was a periodic pulse train near a wavelength of 1550 nm, and the first modulator was not used to remove the onus of spectral alignment of the two microring modulators. The mean photon numbers for the two intensities were about 0.094 and 0.029.

\begin{figure}[!t]
\includegraphics[trim= 0.2in 0.2in 0in 0in, clip, width=3.5in]{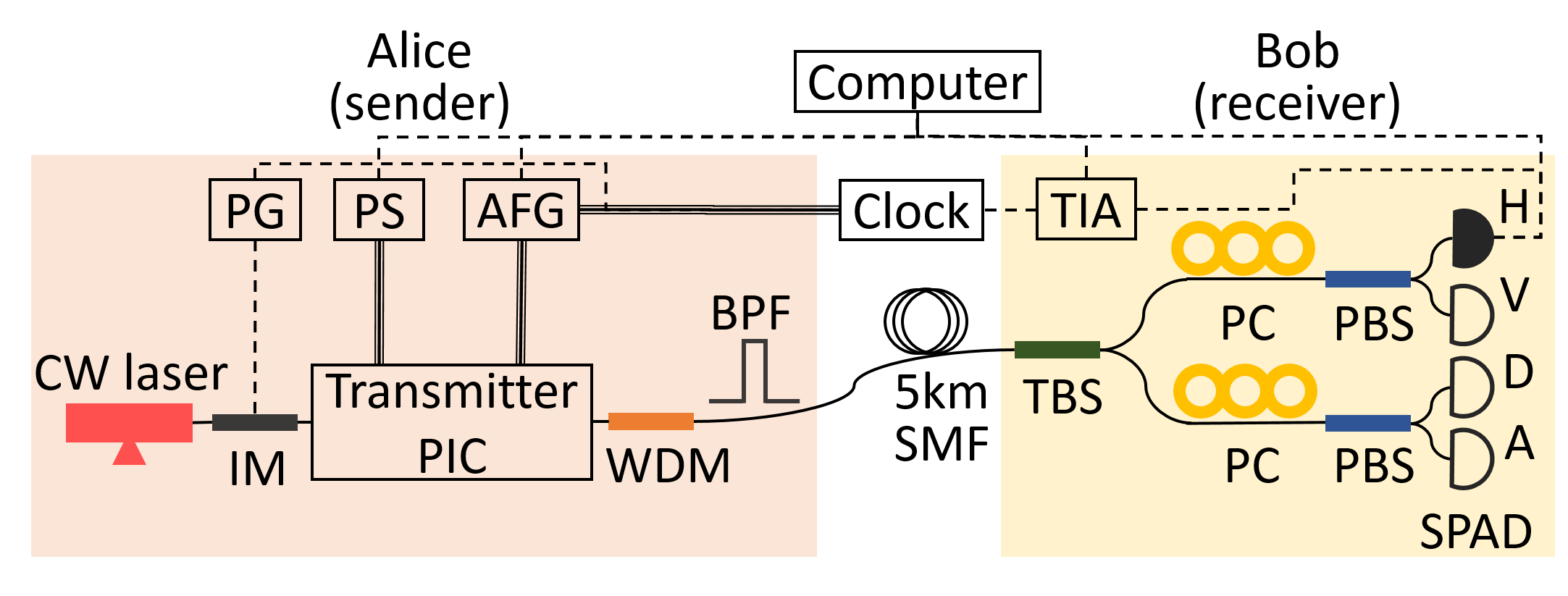}
\caption{\label{fig:setup}Setup schematic. PS: power supply, PG: pulse generator, IM: intensity modulator, AFG: arbitrary function generator, BPF: bandpass filter, SMF: single mode fiber, TBS: tunable beam splitter, PC: polarization controller, PBS: polarization beam splitter, SPAD: single photon avalanche photodetector, TIA: time interval analyzer.}
\end{figure}

%%%%%%%%%%%%%%%%%%%%%%%%%%%%%%%%%%%%%%% 
% QUANTUM KEY DISTRIBUTION
%%%%%%%%%%%%%%%%%%%%%%%%%%%%%%%%%%%%%%%
Finaly,  we present a proof-of-concept BB84 QKD demonstration using the on-chip VOA and  polarization modulator at a wavelength of 1550 nm. The setup is shown in Fig.~\ref{fig:setup}.  For improved modulation ER and thermal stability, the microrings were bypassed and an external LiNbO$_{3}$ intensity modulator was used instead. The external modulator provided a dynamic ER of 30 dB compared to 20 dB of the microring, reducing the penalty on the QBER. For a perfect single photon source, to obtain a secret key with Shor-Preskill's proof \cite{shor2000simple} requires a QBER $\lesssim$11\%. Secure QKD can be done with weak coherent pulses \cite{gottesman2004security}. At the sender (Alice's) side, the input to the  PIC was bursts  of 1000 optical pulses with a FWHM of 1 ns at a repetition rate of 10 MHz. The bursts were at the clock frequency of 9.71 kHz. The PIC attenuated pulses to single photon level and randomly prepared the four polarization states. A wavelength division multiplexer (WDM)  and a bandpass filter (BPF) were added at the output of the PIC to filter out the weak  electroluminesence peak a wavelength of 1150 nm from the forward biased Si PIN diodes \cite{fuyuki2005photographic}. The WDM and BPF added a loss of 5 dB and can be replaced by an integrated on-chip filter in future designs.

The signals were  transmitted over a 5 km long spool of standard single mode fiber. At the receiver (Bob's) side, a tunable beam splitter (TBS) balanced the losses of two paths so that each basis has a 50\% probability to be chosen for the measurement. The PC preceding each PBS was tuned for the measurement of two conjugated bases in the two paths, and the photons were detected using InGaAs single photon avalanche photodetectors (SPADs). The detection efficiency of the SPADs was  20\% and the dead time was set to 15 $\mu$s. The four channels at  Bob's end were measured sequentially, rather than simultaneously, to maintain an identical detection efficiency.  The loss of the  link, which includes the fiber and Bob's equipment, was about 6.1 dB.  The time interval analyzer (TIA) was synchronized with the clock to perform time-correlated single photon counting. 

The on-chip VOA attenuated the pulse intensity by 27 dB. 562 Mbits (effective) were emitted for each channel test. The photon fluxes for the four polarization states had a variation of circa 1 dB, after accounting for the differences in connector loss, in agreement with the classical results. The mean photon per pulse is estimated to be 0.024 and QBER 5.4\%. The raw rate was 13.2 kbps and the asymptotic secure key rate was about 0.95 kbps, which was estimated using the key rate formula without decoy states (see Eq.(12) in \cite{lo2005decoy}). These results show the feasibility of foundry Si photonics for practical QKD, and the key rate would be improved with faster detectors and higher modulation speed.

In summary, we have demonstrated the first Si PIC transmitter for polarization-encoded QKD. Proof-of-concept QKD demonstration showed a QBER of 5.4\% and asymptotic secure key rate of 0.95 kbps. Future improvements include feedback and temperature control for chip stability, PDL compensation, and on-chip integration of the receiver. This work shows the potential of using foundry Si photonics for low cost, wafer-scale manufactured components for optical quantum information. 

\section*{Acknowledgments}
The fabrication was sponsored by CMC Microsystems. We thank Prof. L. Qian for the loan of the SPADs. Funding from the NSERC, Canada Foundation for Innovation, Ontario Research Fund, and Canada Research Chairs is gratefully acknowledged.

\end{document}